\documentclass{aa}

\usepackage{graphicx}
\usepackage{float}
\usepackage{sidecap} 
\usepackage{txfonts}
\usepackage{color}

\usepackage{pifont}

\usepackage{enumitem}
\usepackage{ulem}

\usepackage{color}

\usepackage{tikz}

\usepackage{subcaption}

\usepackage{mhchem}

\usepackage[colorlinks=true,allcolors=blue]{hyperref}

\newcommand{\nH}{$n_{\rm{H}}$}

\newcommand{\mum}{$\mu$m~}
\newcommand{\water}{H$_2$O~}

\newcommand{\BT}[1]{\textcolor{black}{#1}}

\newcommand{\resub}[1]{\textcolor{black}{#1}}

\defcitealias{2021A&A...650A.192T}{T21} 	

\begin{document} 

   \title{OH mid-infrared emission as a diagnostic of H$_2$O UV photodissociation}
  \subtitle{III. Application to planet-forming disks}
  
  \titlerunning{OH mid-infrared emission as a diagnostic of H$_2$O UV photodissociation. III. Planet-forming disks}
  
  \authorrunning{B. Tabone et al.}
  
   \author{Benoît Tabone\inst{1,2}, 
   Ewine F. van Dishoeck\inst{2,3}, 
   John H. Black\inst{4} 
          }
   \institute{      Université Paris-Saclay, CNRS, Institut d’Astrophysique Spatiale, 91405 Orsay, France
      \and
 Leiden Observatory, Leiden University, PO Box 9513, 2300 RA Leiden, The Netherlands
      \and
    Max-Planck-Institut für Extraterrestrische Physik, Giessenbachstrasse1, 85748 Garching, Germany
    \and
    Department of Space, Earth and Environment, Chalmers University of Technology, Onsala Space Observatory, 43992, Onsala, Sweden        
            }    
    
\date{\today}
\abstract
{JWST MIRI gives a unique access to the physical and chemical structure of inner disks ($<10$~au), where the majority of the planets are forming. However, the interpretation of mid-infrared (mid-IR) spectra requires detailed thermo-chemical models able to provide synthetic spectra readily comparable to spectroscopic observations. This is particularly key for OH, which can be excited by a number of processes.}
{Our goal is to explore the potential of mid-IR emission of OH to probe H$_2$O photodissociation, and thus implicitly the far-ultraviolet (FUV) radiation field in the inner disks.}
{We include in the DALI disk model prompt emission of OH following photodissociation of \water in its $\tilde{B}$ electronic state by photons at $\lambda < 144$~nm. \BT{Compared with previous modeling work, we also take into account the propensity of forming OH in the A' symmetric states.} This model allows to compute in a self-consistent manner the thermal and chemical structure of the disk and the resulting mid-IR line intensities of OH and H$_2$O.}
{The OH line intensities in the $9-13~\mu$m range are proportional to the total amount of water photodissociated in the disk. As such, these OH lines are a sensitive tracer of the amount of \water exposed to the FUV field, which depends on the temperature, density, and strength of the FUV field reaching the upper molecular layers. In particular, we show that the OH line fluxes primarily scale with the FUV field emitted by the central star in contrast with \water lines in the 10-20$~\mu$m range which scale with the bolometric luminosity. OH is therefore a key diagnostic to probe the effect of Ly$\alpha$ and constrain the 
dust FUV opacity in upper molecular layers. \BT{A strong asymmetry between the A' and A'' components of each rotational quadruplet is predicted.} 
}
{OH mid-IR emission is a powerful tool to probe \water photodissociation and infer the physical conditions in disk atmospheres. As such, the inclusion of OH mid-IR lines in the analysis of JWST-MIRI spectra will be key for robustly inferring the chemical composition of planet-forming disks. 
The interpretation of less excited OH lines in the MIRI-MRS range requires additional quantum calculations of the formation pumping of OH (ro-)vibrational levels by O+H$_2$ and the collisional rate coefficients.}           


\maketitle


\section{Introduction}
Planets form, migrate, and acquire their elemental composition in disks orbiting nascent stars. In the next decade, our knowledge of planetary systems is expected to grow considerably, with a more complete view of the frequency and distribution of planets (PLATO), but also with the determination of the elemental abundances in their atmosphere (JWST, ARIEL). One of the challenges is then to determine how the various formation pathways lead to the diversity and habitability of exoplanets.

In this context, the atmospheric elemental composition of gas giants is expected to carry key information about their formation \citep{2019ARA&A..57..617M}. In gas-rich disks, variation in the elemental composition in the gas and in solids (ice and dust grains) is a natural outcome of the different sublimation temperatures of the main carriers of elements. This led to the idea proposed by \citet{2011ApJ...743L..16O} that the final elemental composition of planets is set by the location where they formed \citep[see also][]{2019A&A...627A.127C,2019A&A...632A..63C,2020A&A...642A.229C}. Since then, many other complex processes that can modify the elemental composition in disks have been identified \citep{2022arXiv220310056K} such as trapping molecules in the cold mid-plane \citep[a.k.a. "cold finger effect",][]{2009ApJ...704.1471M}, drifting icy pebbles \citep{2016ApJ...833..285K}, or destruction of carbon dust \citep{2017A&A...606A..16G}. 

Of all the species, water is a unique molecule for planet formation. Together with CO, it is the main oxygen carrier and as a result, the location of the water snowline sets the region where oxygen-rich planetesimals form and where the gas is depleted in oxygen. Water ice can also enhance the sticking of grains, allowing planetesimals to grow fast at the ice line and initiate the formation of the cores of giant planets or super-Earths \citep{1993ApJ...407..806C,2017A&A...602A..21S}. Gas-phase water plays also a crucial role in the irradiated upper layers where most of the IR emission to be observed by JWST originates. Its 
emission is prominent in the infrared spectra of T Tauri disks \citep{2008Sci...319.1504C,2008ApJ...676L..49S} \resub{even in depleted inner disks like DoAr 44 \citep{2015ApJ...810L..24S} or PDS 70 \citep{2023Natur.620..516P}}. Its numerous radiative transitions contribute to the thermal balance and water photodissociation can efficiently heat the gas \citep{2015ApJ...810..125G}. Gas-phase water in the irradiated layers can also shield the disk by absorbing dissociating UV photons \citep{2009Sci...326.1675B,
2014ApJ...786..135A,2022ApJ...930L..26B}. This process can greatly alter the abundances of other key species such as CO$_2$ \citep{2022ApJ...933L..40B} or hydrocarbons \citep{2022ApJ...934L..25D}. These latter works demonstrate that variation in the abundances of C- or O-bearing species cannot simply be associated with variation in the C/O ratio but can also be due to the ability of an H$_2$O-rich irradiated layer to shield the disk. Therefore the study of \water and its photodissociation in the irradiated layers is of fundamental interest to understanding the physical and chemical structure of disks.

Considering the complexity of processes that alter the distribution of chemical elements in disks, robust observational constraints are needed. ALMA has brought unique insights into disk chemistry \citep{2014prpl.conf..317D,2021PhR...893....1O}. However, the majority of gaseous planets that are about to be routinely characterized by JWST or ARIEL are thought to form in inner disks \citep[$\lesssim 10~$au,][]{2016JGRE..121.1962M,2008ApJ...673..502K,2018ARA&A..56..175D}, best traced by near- and mid-infrared (IR) molecular lines. These regions were made accessible thanks to the \textit{Spitzer} space telescope supplemented by ground-based observations \citep[VLT/CRIRES, TEXES, Michelle][]{2014prpl.conf..363P}. A rich organic chemistry has indeed been unveiled with the detections of the main carbon and oxygen carriers, namely H$_2$O, CO, CO$_2$, HCN \citep{2010ApJ...720..887P}. Slab models and spectrally resolved ground-based observations further showed that IR emission in the 5-36 \mum range originates from warm gas (500-1000 K) within a few au \citep{2011ApJ...731..130S,2011ApJ...733..102C}. The observed spectra
of T Tauri stars show also significant diversity \citep{2009ApJ...696..143P,2013ApJ...779..178P} that could be driven by a variation in the C/O ratio \citep{2013ApJ...766..134N,2020ApJ...903..124B} or inner cavities \citep{2017ApJ...834..152B}. Therefore, the mid-IR constitutes a unique domain as it probes the gas in the planet-forming regions of disks, where gas-phase \water plays a crucial role. Today, the James Webb Space Telescope (JWST), with its unique sensitivity and good spectral resolution, is revolutionizing the field \citep{2023arXiv230711817V} by detecting weaker lines and isotopologues from previously detected species \citep{2023ApJ...947L...6G} and new molecular species \citep{2023NatAs...7..805T,2023Natur.621...56B}.

However, the interpretation of mid-IR molecular emission requires detailed modelling. Thermochemical models show that the IR emission comes from different layers, depending on the ability of each species to survive strong UV fields and on physical conditions required to excite the lines \citep[see e.g.,][]{2018A&A...618A..57W}. Therefore the ratio of column densities obtained from slab models are difficult to interpret and detailed thermochemical models are ultimately needed to retrieve the physical and chemical structure of disks from the rich IR spectra.

Up to now, there has been a significant effort to model \water chemistry and its IR emission in disks. The pioneering work of \citet{2009ApJ...704.1471M} showed that water line fluxes observed by \textit{Spitzer}-IRS hint at a significant depletion of small grains in disk atmospheres. The main limitation of all these models is that they are based on \water lines that are excited by collisions and IR radiative pumping \citep{2016ApJ...818...22B}. As such, their intensity depends on a complex combination of the density, temperature, and IR local radiation fields across the disk. Analysis relying on only water lines remains therefore degenerate \citep{2015A&A...582A.105A}.

In contrast, rotationally excited lines of OH lying in the mid-IR are expected to provide complementary constraints on water in IR active layers. Water photodissociation in its $\tilde{B}$ electronic state by photons in the 114-144~nm range produces OH in highly rotationally excited states (up to $N \simeq 45$, corresponding to $E_{\text{up}} \simeq 40,000$~K). The subsequent radiative deexcitation of OH products leads to a series of rotationally excited lines longward of $9~\mu$m, a process called prompt emission. OH lines from highly excited rotational levels were first detected at the apex of a fast protostellar jet \citep{2008ApJ...680L.117T} and later in disks \citep{2010ApJ...712..274N,2014ApJ...788...66C}. Thanks to previous quantum calculations and experiments \citep{2000JChPh.11310073H,2000JChPh.112.5787V}, these authors already identified this emission as prompt emission due to H$_2$O phtodissociation. \citet{2011ApJ...733..102C} showed a correlation between the accretion rate and the flux of rotationally excited lines, further highlighting the role of the UV field in the excitation of these lines. In disks, less excited rotational lines longward of $16-20~$\mum are often attributed to thermal excitation \resub{or chemical formation pumping by H$_2$+O \citep{2014ApJ...788...66C}.} \resub{The signature of H$_2$O photodissociation in its $\tilde{B}$ state is also likely seen in the FUV where a bump at 160~nm is attributed to dissociating H$_2$ following its production by water photodissociation \citep{2017ApJ...844..169F}. Finally, water photodissociation in the $\tilde{A}$ state produces rotationally cold but vibrationally hot OH \citep{2001JChPh.114.9453V} which produces ro-vibrational emission around 3~$\mu m$. Yet, formation pumping by H$_2$+O or collisions are likely to dominate the excitation of the ro-vibrational levels of OH in disks, as recently found in the wind of an externally irradiated disk \citep{2024NatAs.tmp...47Z}.}

In the first paper of this series, we demonstrated that the fluxes of rotationally excited lines of OH are directly proportional to the amount of water photodissociated per unit time in the 114-144~nm range \citep{2021A&A...650A.192T} (\citetalias{2021A&A...650A.192T}, hereafter). In a slab approach (single-zone) and without relying on chemical models, the local UV flux can be inferred from \water and OH mid-IR lines. However, this approach is limited to simple geometries and uniform physical conditions (at least within the beam of the telescope) which is rarely the case in irradiated environments such as interstellar PDRs or disk upper layers. On the other hand, when relying on a self-consistent thermo-chemical model, the OH line flux can be turned into a sensitive diagnostic. This has been demonstrated in the second paper of this series which shows that in interstellar PDRs, the OH line flux is highly sensitive to the thermal pressure of the gas \citep{2023A&A...671A..41Z}. 

In this paper, we expand our modeling effort to inner disks to explore the potential of OH mid-IR lines to trace H$_2$O photodissociation and to alleviate degeneracies in the analysis of molecular lines. In particular, the excitation of OH following H$_2$O photodissociation via the $\tilde{B}$ electronic state is included in the DALI physico-chemical disk model. This allows us to connect the mid-IR emission of OH to the physical and chemical structure of the disks' upper layers. This paper is organized as follows. In Sec. \ref{sec:model} we present the basics of DALI, the implementation of OH prompt emission, and the free parameters of the model. The processes that control OH emission are detailed using our fiducial disk model and the result of a grid of models are described in Sec. \ref{sec:results}. MIRI-MRS predictions are finally presented based on previous \textit{Spitzer}-IRS observations of T Tauri disks in Sec. \ref{sec:discussion}. Our findings are summarized in Sec. \ref{sec:conclusion}.

\section{Model}
\label{sec:model}

\begin{table}
\caption{Parameters of the DALI model}              
\label{table:grid-parameters}      
\centering                                      
\begin{tabular}{c c c c c}          
\hline\hline                        
Parameter &   & Fiducial & Range \\   
 \hline\hline 
Mass       & $M_* [M_{\odot}]$ & 1.0 & . \\
\textbf{Luminosity} & $L_* [L_{\odot}]$ & 1 & 0.5-4.5\\
Effective temperature & $T_{eff} [K]$  & 4250 & .\\
\textbf{Accretion luminosity} & $L_{acc} [L_{\odot}]$ & 0.12 & 0.012-0.4 \\
\textbf{Ly-a contribution} &  wrt. $L_{acc}$ &  $0$ & $0$ and $0.15$ \\
 \hline\hline 
Disk mass & $M_{D} [M_{\odot}]$ &  0.03 & . \\
Disk size & $R_{c} [au]$ &  46 & . \\
Disk aspect ratio & $h_{c}$ &  0.2 & . \\
Flaring angle & $\psi$ &  0.11 & . \\
\textbf{Gas-to-dust mass ratio} &  $d_{gd}$ &  $10^{5}$ & $10^{2}-10^{5}$ \\
\hline   
\end{tabular}
\end{table}

\subsection{Thermo-chemical disk model}

In this work, the intensities of OH and H$_2$O lines are computed in a self-consistent approach using an upgraded version of the 2D thermo-chemical code DALI \citep{2012A&A...541A..91B,2022ApJ...930L..26B}. This code has been extensively used to analyze multi-wavelength observations of disks, from the millimeter to infrared domain \citep[e.g.,][]{
2013A&A...559A..46B,
2018A&A...609A..93C,
2019A&A...629A..79T,2019A&A...631A.133B, 2021A&A...646A...3L}, and is particularly well suited to model inner disk atmospheres \citep{2015A&A...575A..94B,2018A&A...611A..80B,2019A&A...631A.133B, 2022ApJ...933L..40B}. Given an input gas and dust density structure, the temperature of the dust and the local radiation field are computed via a 2D Monte Carlo method. The thermo-chemical state of the gas is then obtained following an iterative two-step procedure in each grid cell: first, the chemical abundances are computed assuming steady-state, and then the gas temperature is computed from the balance between heating and cooling \citep{2012A&A...541A..91B}. The latter step includes radiative cooling via atomic and molecular lines which involves solving the non-LTE excitation of a number of key species \citep{2013A&A...559A..46B}. The emerging line intensities are eventually computed using a fast ray-tracer method \citep{
2017A&A...601A..36B} assuming a distance of 140~pc and an inclination of 20$^{\circ}$. 

The key chemical reactions for oxygen chemistry in the warm molecular layers involve the formation of \water via the warm route
\begin{equation}
    \ce{O <=>[H_2][UV] OH <=>[H_2][UV] H_2O}.
    \label{eq:O-chemistry}
\end{equation}
The sequence of reaction requires the presence of H$_2$. The latter species can form via (i) formation on dust grains, for which the rates stem from \citet{2004ApJ...604..222C}, (ii) the three-body formation route, (iii) formation via CH$^+$+H and H$^-$+H following radiative associations \citep[see e.g.,][]{2020A&A...636A..60T}, and (iv) formation via PAH hydrogenation. All these routes are included, following the implementation of \citet{2022ApJ...930L..26B} for the former two, and \citet{2012A&A...541A..91B} for the latter two.

In order to improve the treatment of oxygen chemistry in the warm upper layers of inner disks, we adopt the DALI version from \citet{2022ApJ...930L..26B} that includes chemical heating following H$_2$O, OH, CO, H$_2$, and C photodestruction \citep{2015ApJ...810..125G}, and the wavelength-dependent UV shielding of the gas by H$_2$O. Since our models are designed to describe the warm and dense atmospheres of inner disks, the cooling via rovibrational lines of CO ($\varv \le 4$) and H$_2$O is included using the collisional rate coefficients from \citet{2010ApJ...718.1062Y,2015ApJ...811...27W,2015ApJ...813...96S} and \citet{2008A&A...492..257F}, respectively. Cooling by OH lines is also included as described in the following section.

\subsection{OH model}

The excitation of OH is solved in concert with the excitation of the other species. In order to include prompt emission, we assume that only H$_2$O photodissociation leads to the production of OH in super-thermal state distribution and that the destruction rate of an OH molecule does not depend on its quantum state. This leads to the detailed balance equation
\begin{equation}
\frac{d n_i}{d t} = \sum_{j \neq i} P_{ji} n_j - n_i \sum_{j \neq i} P_{ij} +  F_{pd} \left( \bar{f_i} - \frac{n_i}{n(\text{OH})} \right) = 0,
\label{eq:statistical-eq}
\end{equation}
where $n_i$ $[\rm{cm}^{-3}]$ is the local population densities of OH, $F_{pd}$ is the production rate of OH via H$_2$O  photodissociation $[\text{cm}^{-3} \rm{s}^{-1}]$, and $\bar{f_i}$ is the probability to form OH in the state $i$ following H$_2$O photodissociation by the local UV radiation field. $P_{ij}$ are the radiative and collisional transition probabilities $i \rightarrow j$ given by
\begin{equation}
  P_{ij}=\begin{cases}
     A_{ij} + B_{ij} \bar{J}_{\nu_{ij}}+ C_{ij}  & (E_i > E_j)\\
     B_{ij} \bar{J}_{\nu_{ij}} + C_{ij} & (E_i < E_j).
  \end{cases}
  \label{eq:Pij}
\end{equation}
$A_{ij}$ and $B_{ij}$ are the Einstein coefficients of spontaneous and induced emission, $C_{ij}$ are the collisional rate coefficients, and $\bar{J}_{\nu_{ij}}$ is the mean specific intensity at the frequency of the radiative transition $i \rightarrow j$ averaged over the line profile. Following \citet{2013A&A...559A..46B}, the  contribution of the lines to the local radiation field is computed in a 1+1D approach only accounting for the vertical and radial direction to obtain the escape probabilities.

The list of OH levels and radiative transitions stem from \citetalias{2021A&A...650A.192T} who used data from \citet{2018JQSRT.217..416Y} and \citet{2016JQSRT.168..142B}. In order to reduce the computational time, the number of OH levels has been reduced to a total of 412 by limiting the vibrational quantum number to $\varv \le 1$ and including only the OH($\tilde{X}$) electronic ground state. All the rotational levels that are stable within a vibrational state are retained, which corresponds to $N \le 50$ and $N \le 48$ for $\varv = 0$ and $1$, respectively. Each rotational level is further split by the spin-orbit coupling labeled by $\Omega =1/2,3/2$ and the $\Lambda$-doubling labeled by the $e/f$ parity. Throughout this paper, we also use the A' and A'' symmetry (with respect to reflection about the plane of rotation of the molecule) to designate states with $\Omega =1/2, f$ or $\Omega =3/2, e$ and $\Omega =1/2, e$ or $\Omega =3/2, f$, respectively. As in \citetalias{2021A&A...650A.192T}, we consider intra- and cross-ladder rotational transitions in the $\varv=0$ and $\varv=1$ bands as well as in between the $\varv=1$ and $\varv=0$ states, resulting in a total of 2360 (ro-)vibrational transitions.

The distribution of nascent OH denoted as $\bar{f_i}$ in Eq. (\ref{eq:statistical-eq}) stems from the compilation of \citetalias{2021A&A...650A.192T}. In this work, some of the OH ro-vibrational levels that have been discarded can be populated by prompt emission. In practice, the radiative cascade from these levels can impact the population of the rotational lines in the ground vibrational state. Therefore, a reduced distribution of nascent OH has been calculated assuming that all the OH products formed in levels that have been discarded do cascade toward the set of retained levels. This assumption is valid up to large densities (\nH $\lesssim 10^{12}$ cm$^{-3}$) and strong IR radiation fields. \BT{Following the recent finding of \citet{2024NatAs.tmp...47Z} and in line with recent quantum dynamical calculations \citep{2015JChPh.142l4317Z}, we also assume that H$_2$O photodissociation produces OH exclusively in the rotational states with an $A'$ symmetry $\Omega =1/2, f$ and $\Omega =3/2, e$}.

The exact distribution of nascent OH depends on the energy of the FUV photon. However, the main difference in the distribution is between photodissociation via the H$_2$O($\tilde{A}$) state, long-ward of 145nm, that produces OH in vibrationally hot but rotationally cold states \citep{2001JChPh.114.9453V}, and photodissociation via the H$_2$O($\tilde{B}$) state, shortward of 145nm, that produces OH in rotationally hot states \citep{2003CPL...370..706V}. In this work, we focus on the mid-IR lines of OH that are rotationally excited and therefore impacted by photodissociation via the $\tilde{B}$ state and not the $\tilde{A}$ state. We further neglect the wavelength dependency of the distribution of nascent OH following photodissociation through the $\tilde{B}$ band and adopt Ly $\alpha$ ($\lambda=121.6~$nm) as a representative wavelength to compute the distribution of nascent OH. Formation pumping by the O+H$_2 \rightarrow$ OH($v,N$) + H reaction is not included in order to highlight the effect of prompt emission. We also stress that chemical pumping impacts relatively low-lying rotational levels of OH traced by lines longward of 16 $\mu$m.

The collisional rate coefficients for He and H$_2$ stem from the quantum calculations of \citet{2007CPL...445...12K} and van den Heuvel et al. (in prep.). \resub{They have been further extrapolated assuming $k \propto e^{-\Delta E/k_bT}$. Regarding collisions with He, the collisional rate coefficients for the A'/A'' changing transitions within a $N$ level are poorly fitted by the latter ansatz. For these transitions, we therefore adopt the rates of the highest computed $N$ transition of \citet{2007CPL...445...12K}.} \resub{In the absence of data, collisions with atomic hydrogen are not included in our model.}


\begin{figure*}[!t]
\centering
\includegraphics[width=.9\textwidth]{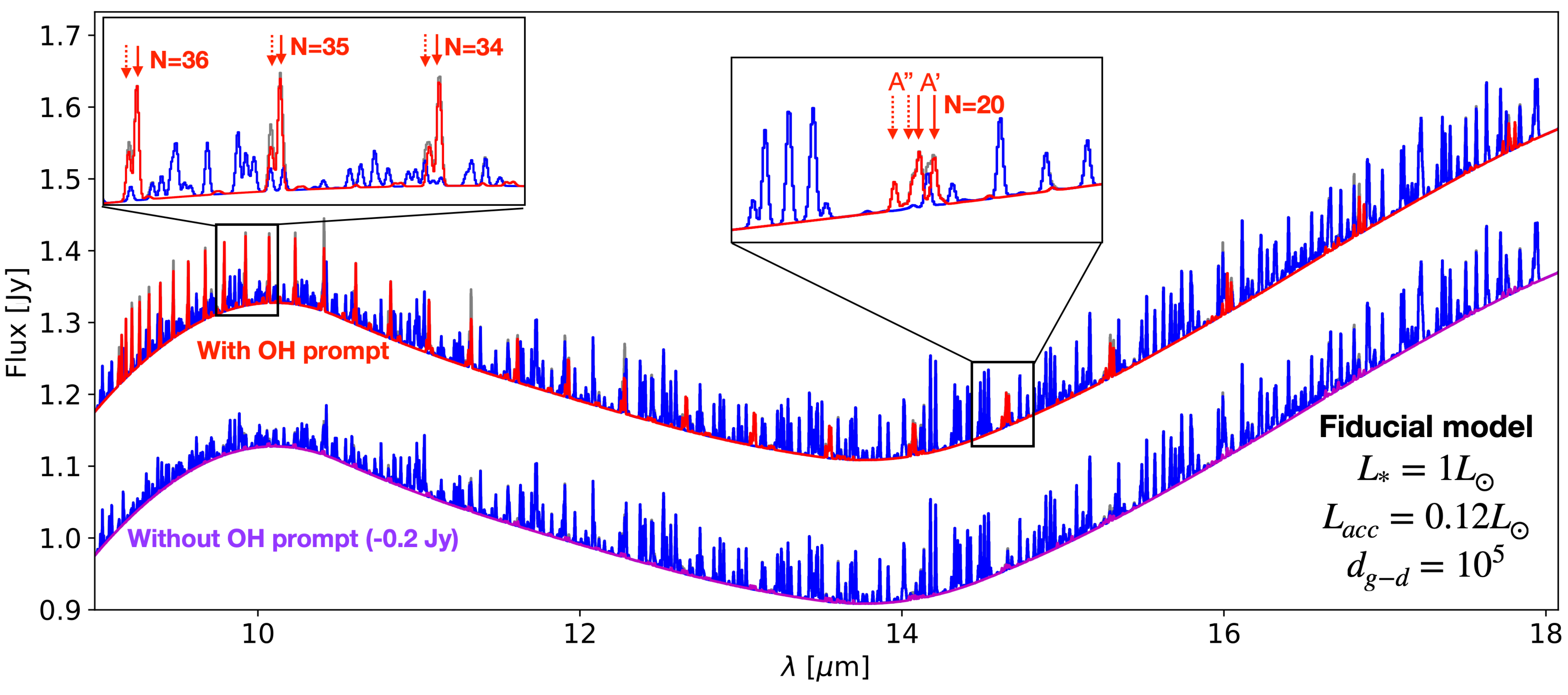}
\caption{Synthetic spectrum of OH (red and purple) and H$_2$O (blue) for the fiducial model, at typical JWST-MIRI spectral resolution ($R=2500$) with (top) and without (bottom, shifted by 0.2 Jy) OH prompt emission. OH is primarily excited by H$_2$O photodissociation producing highly excited lines down to 9~$\mu$m. The OH lines are seen on top of the continuum and embedded in the strong H$_2$O line forest. Note that OH prompt emission produces OH in the $A'$ symmetric states resulting in an imbalance in the four components of each rotational transition. Longward of $\simeq 13.5~\mu$m, we expect OH chemical pumping by O+H$_2$ to impact the OH lines, a process that is not included in this version of our model.
}
\label{fig:spectrum-fiducial}
\end{figure*}

\subsection{Setup and free parameters}

Our goal is to explore how the mid-IR emission of OH and \water relates to the stellar properties and to the distribution of gas and dust in the disk. Therefore, we do not model the evolution of gas and dust (e.g., hydrostatic equilibrium, dust settling, and radial drift) but we use instead a parameterized disk model. This allows us to obtain predictions that do not depend on sophisticated assumptions and theoretical preconceptions. The parameters and their values explored in this work are reported in Table \ref{table:grid-parameters}.

The surface density profile of the gas is
\begin{equation}
\Sigma_{gas}(R) = \frac{M_D}{2 \pi R_c^2} \left( \frac{R}{R_c} \right)^{-1} \exp{\left( -\frac{R}{R_c} \right)} ,
\end{equation}
where the characteristic disk radius is set to $R_c = 46~$au and the disk mass to $M_D = 0.03 M_{\odot}$. The vertical distribution of the gas follows an effective isothermal profile of
\begin{equation}
\rho_{gas}(\theta, R) = \frac{\Sigma_{gas}(R)}{\sqrt{2 \pi} R h(R)} \exp{\left( -\frac{\pi/2 - \theta}{2 h(R)} \right)},
\end{equation}
with a disk aspect ratio of $h(R) = h_0 (R/R_c)^{\psi}$. 

Regarding the dust, we consider only the "small dust" population used by \citet{2013A&A...559A..46B}, which corresponds to a mixture of silicate and graphite with a size distribution ranging from 5~nm to 1~$\mu$m. The grains are also assumed to be fully mixed with the gas, with a homogeneous gas-to-dust mass ratio denoted as $d_{gd}$, that is considered as a free parameter, with a fiducial value of $d_{gd}=10^{5}$, in line with \textit{Spitzer} observations \citep{2006ApJS..165..568F}. Our model does not include a population of large grains because those grains are settled in the mid-plane and have a minimal impact on the IR lines emitted from the upper layers. The PAH abundance is set to $10^{-6}$ (wrt. ISM abundance) which gives a negligible contribution of PAH on the disk thermochemistry.


The disk is assumed to be irradiated by a T Tauri star with an effective temperature of $T_{eff} = 4,250 K$ and a photospheric luminosity $L_*$. \resub{The exact value of the effective temperature has little impact on the resulting thermochemical structure as long as its FUV luminosity is negligible compared to the UV excess due to accretion which is the case for accreting low-mass stars T Tauri stars ($T_{eff} \lesssim 5,000~K$).} The stellar X-ray spectrum corresponds to a blackbody at $T_X=4.6 \times 10^7~$K between $10^3$ and $10^5$eV with a luminosity of $L_X=10^{30}$ erg s$^{-1}$. \BT{The FUV continuum excess is modeled as a blackbody at $20,000$~K with a total luminosity of $0.14~L_{acc}$, corresponding to a FUV luminosity (integrated between 116-170~nm) of $L_{FUV,cont} = 0.03 L_{acc}$, in line with the FUV continuum excess measured in T~Tauri disks \citep{2012ApJ...756L..23S,2014ApJ...784..127F}. We stress that the ratio between the FUV luminosity and the accretion luminosity varies significantly from source to source. In this work, we use the accretion luminosity as a proxy of  $L_{\rm{FUV,cont}}$  because $L_{acc}$ is more systematically measured in a large sample of stars. Still, one should keep in mind that when comparing our results to observations, one should prefer to use the FUV luminosity whenever available.}

FUV observations show that Ly$\alpha$ photons contribute to about 80\% of the total stellar FUV luminosity \citep{2003ApJ...591L.159B,2012ApJ...756L..23S}. However, the amount of  Ly$\alpha$ reaching the IR active molecular layers depends on the scattering by H atoms followed by absorption by dust or gas. The consistent calculation of Ly$\alpha$  propagation is beyond the scope of the paper and in this work, we neglect scattering by H atoms as done in most thermochemical disk models \citep[e.g.,][]{2015A&A...582A..88W,2009A&A...501..383W}. We however explore the impact of Ly$\alpha$ photons in specific models by adding a Ly$\alpha$ line with a FWHM of $200$ km/s and a total luminosity of $0.15~L_{acc}$.


In short, our model is controlled by four free parameters: the stellar luminosity $L_*$, the accretion luminosity $L_{acc}$, the contribution of Ly$\alpha$ photons, and the gas-to-dust mass ratio $d_{g/d}$. We define a fiducial model that corresponds to a line-rich T Tauri disk with $L_* = 1 L_{\odot}$, $L_{acc} = 0.12 L_{\odot}$, and $d_{g/d} = 10^5$, and no contribution of Ly$\alpha$.

\section{Results}

\label{sec:results}

\subsection{Fiducial model}

\label{subsec:result-fiducial}

\subsubsection{Prompt emission spectrum}

Figure \ref{fig:spectrum-fiducial} illustrates the importance of prompt emission in the mid-infrared spectrum of our fiducial disk model. Prompt emission enhances the OH line intensities shortward of $\simeq$25~\mum with an amplifying factor that decreases with decreasing wavelength. As discussed in \citetalias{2021A&A...650A.192T}, \water photodissociation via the $\tilde{B}$ state produces rotationally excited OH. The subsequent radiative cascade leads to an increase in $N \rightarrow N-1$ rotational lines. Compared with thermal excitation (bottom spectrum in Fig. \ref{fig:spectrum-fiducial}), prompt emission will enhance more the rotational lines that are emitted by high rotational levels. This is particularly the case shortward of 20~$\mu$m range, which corresponds to quantum rotational levels of $N>14$. In this specific model, thermal excitation and IR pumping start to take over prompt emission for rotational levels of $N \simeq 11$, corresponding to upper energies of $E_{\text{up}} \simeq 3,000~$K. 
We however stress that for upper energy levels up to $N \simeq 25$ ($\lambda \gtrsim 13~\mu$m), the formation pumping of OH via O+H$_2$ can enhance OH lines. Today, the exact distribution of OH($v,N$) following its formation remains to be calculated with quantum dynamical calculations. In the following, we therefore focus on the lines shortward of 13~$\mu$m which are uniquely excited by water photodissociation.

Each rotational line of OH is split by the spin-orbit interaction and the $\Lambda$-doubling forming a quadruplet. Shortward of $\simeq 12~\mu$m, only two components of the quadruplet are spectrally resolved by MIRI-MRS, forming a doublet of opposite symmetry denoted as A' and A'' (see Sec. \ref{sec:model}). \BT{Figure \ref{fig:spectrum-fiducial} (top left insert) shows that our DALI model predicts a strong asymmetry in between the A' and A'' components. This is a clear signature of water photodissociation producing OH preferentially in the A' states. This effect was not included in previous modelling work of OH prompt emission \citep{2021A&A...650A.192T,2023A&A...671A..41Z}.
Interestingly, the A'' component is still visible in our synthetic spectrum whereas the model assumed OH production exclusively in the A' state. This is due to collisions at the inner edge of the disk, which populate the A'' states from the A' states. However, we stress that collisional de-excitation of the highly excited levels of OH remains uncertain since the collisional rates are extrapolated from the low $N$ states of OH. Dedicated quantum calculations of the collisional rate coefficients are needed to interpret the exact A''/A' ratio that can be measured by JWST-MIRI.}

As shown by \citetalias{2021A&A...650A.192T}, the shape of the prompt emission spectrum of OH in the mid-IR is set by the distribution of nascent OH following \water photodissociation shortward of 145 nm. In our DALI models, this simple property pertains due to the limited effect of collisions. As such, and because of our simple calculation of the distribution of OH products (see Sec. \ref{sec:model}), the relative line intensities are always the same. Therefore, in the following, we focus only on the line at $10.07 \mu$m that is excited enough to be always dominated by prompt emission across the disk ($N=34$, $E_{\text{up}}=28,300~$K$, A_{i,j}= 435~$s$^{-1}$). This line is also less contaminated by \water lines. The flux of other rotationally excited lines can be deduced from that of the $10.07~\mu$m line by the simple rescaling factors provided in \citetalias{2021A&A...650A.192T} (see their Appendix D).

\subsubsection{Emitting area and origin of the OH emission}

\begin{figure}
\centering
\includegraphics[width=0.55\textwidth]{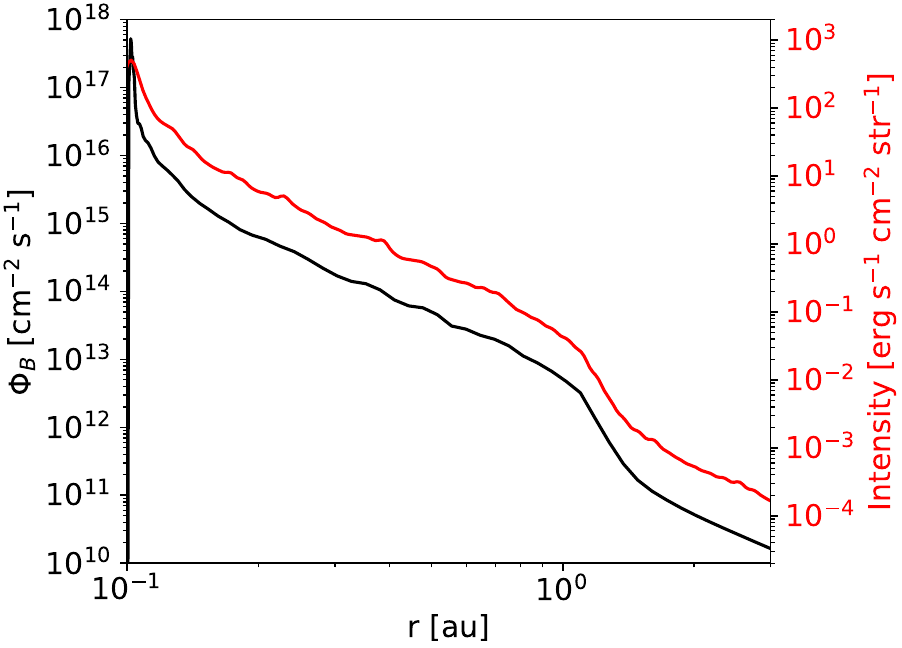}
\caption{Radial profile of the emission of the 10.07 \mum OH line and of the \water column density photodissociated per unit time in the 114-145~nm FUV range. Throughout the disk, the line intensity is proportional to the amount of \water locally photodissociated via the $\tilde{B}$ state, except in the innermost part of the disk where the line gets optically thick and geometrical effects play a role.}
\label{fig:phi-fiducial}
\end{figure}

In \citetalias{2021A&A...650A.192T}, we showed that the intensities of rotationally excited OH lines are proportional to the column density of \water photodissociated per unit time along the line-of-sight defined as
\begin{equation}
   \Phi_B = \int n_{\rm{H_2O}}(z) k_{B}(z) d z~~~\rm{[cm^{-2}~s^{-1}]},
    \label{eq:def_PhiB}
\end{equation} 
where $k_{B}(z)$ is the photodissociation rate of H$_2$O in the $\tilde{B}$ band ($114-144~\rm{nm}$) leading to OH. Figure \ref{fig:phi-fiducial} demonstrates that this simple property pertains across most of the radial extent of the disk. In particular, the specific intensity of the 10.07~\mum OH line scales as
\begin{equation}
    I \simeq 1.1 \times 10^{-4}  \left( \frac{\Phi_B}{10^{10}} \right)~\rm{erg ~s}^{-1}~\rm{cm}^{-2}~\rm{sr}^{-1},
    \label{eq:I_PhiB}
\end{equation} 
in line with the scaling provided in \citetalias{2021A&A...650A.192T}, with the modification that OH is only produced in the A' state.
In the innermost part of the disk, the intensity differs from this scaling. This is due to the inclination of the disk and due to the OH line becoming optically thick above $\Phi_B \simeq 10^{17} \rm{cm^{-2}~s^{-1}}$. \resub{We stress that order of magnitude estimates of the OH-H collisional rate coefficients are needed to check if collisional deexcitation with H can quench prompt emission in the innermost region.}

\begin{figure}
\centering
\includegraphics[width=0.5\textwidth]{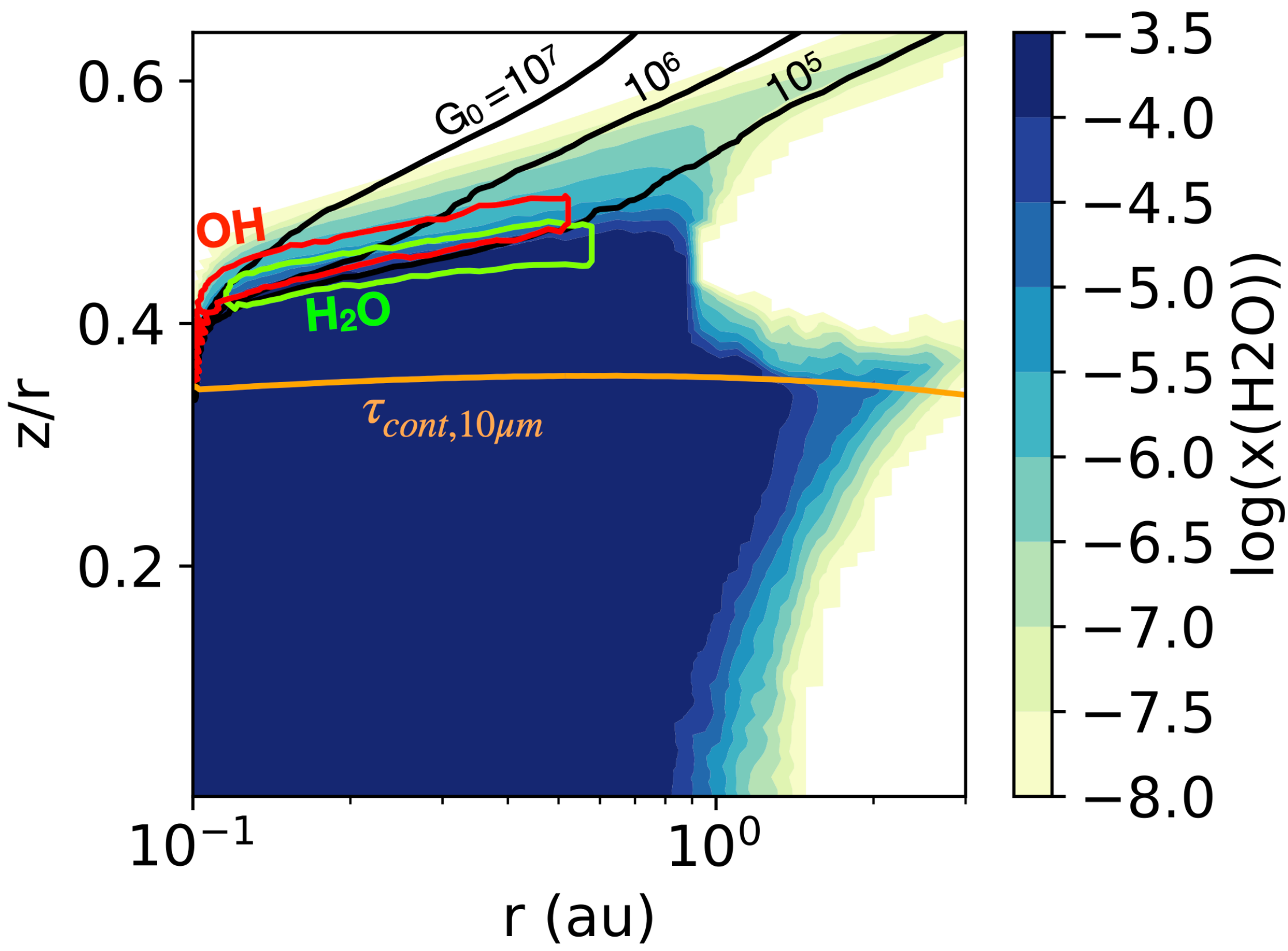}
\caption{Disk thermochemical structure of our fiducial model. The emitting area of the rotationally excited OH line at 10.07~\mum and the \water line at 12.519~\mum are shown on top of the distribution of the gas-phase water abundance in the disk (color map) and the UV radiation field (black contours) for our fiducial model.}
\label{fig:dali-intro-physico-chem}
\end{figure}

\begin{figure}
\centering
\includegraphics[width=0.45\textwidth]{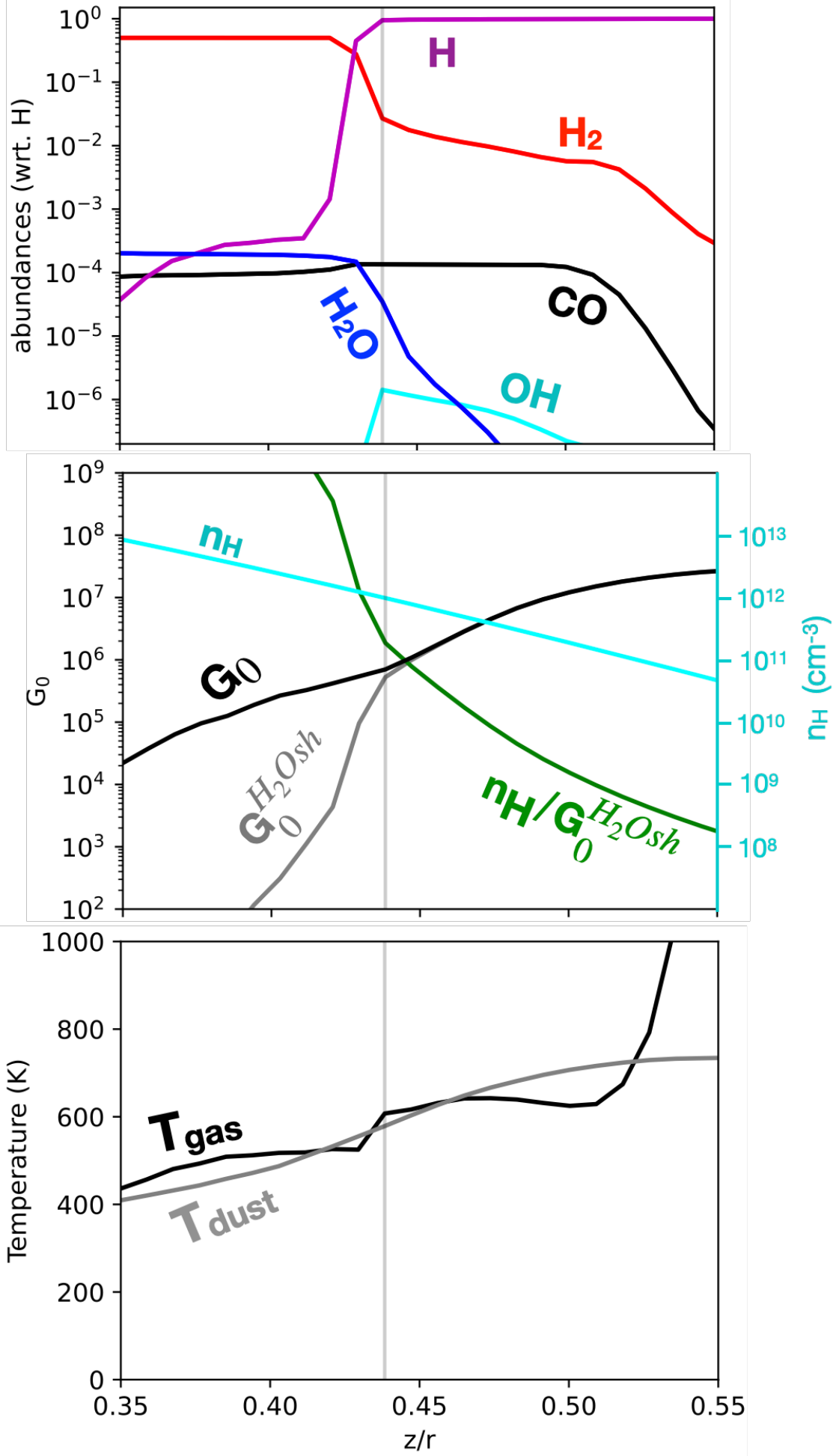}
\caption{Physical and chemical structure of the irradiated layer at $r=0.3$ au for our fiducial disk model. $G_0^{H_2O sh}$ and $G_0$ are the local FUV radiation field attenuated by gas and dust, and by dust only, respectively in \resub{Draine} units. The difference of these two quantities below $z/r \lesssim 0.46$ reveals the UV shielding by \water.}
\label{fig:fiducial-vertical-structure}
\end{figure}

\begin{figure*}
\centering
\includegraphics[width=1.\textwidth]{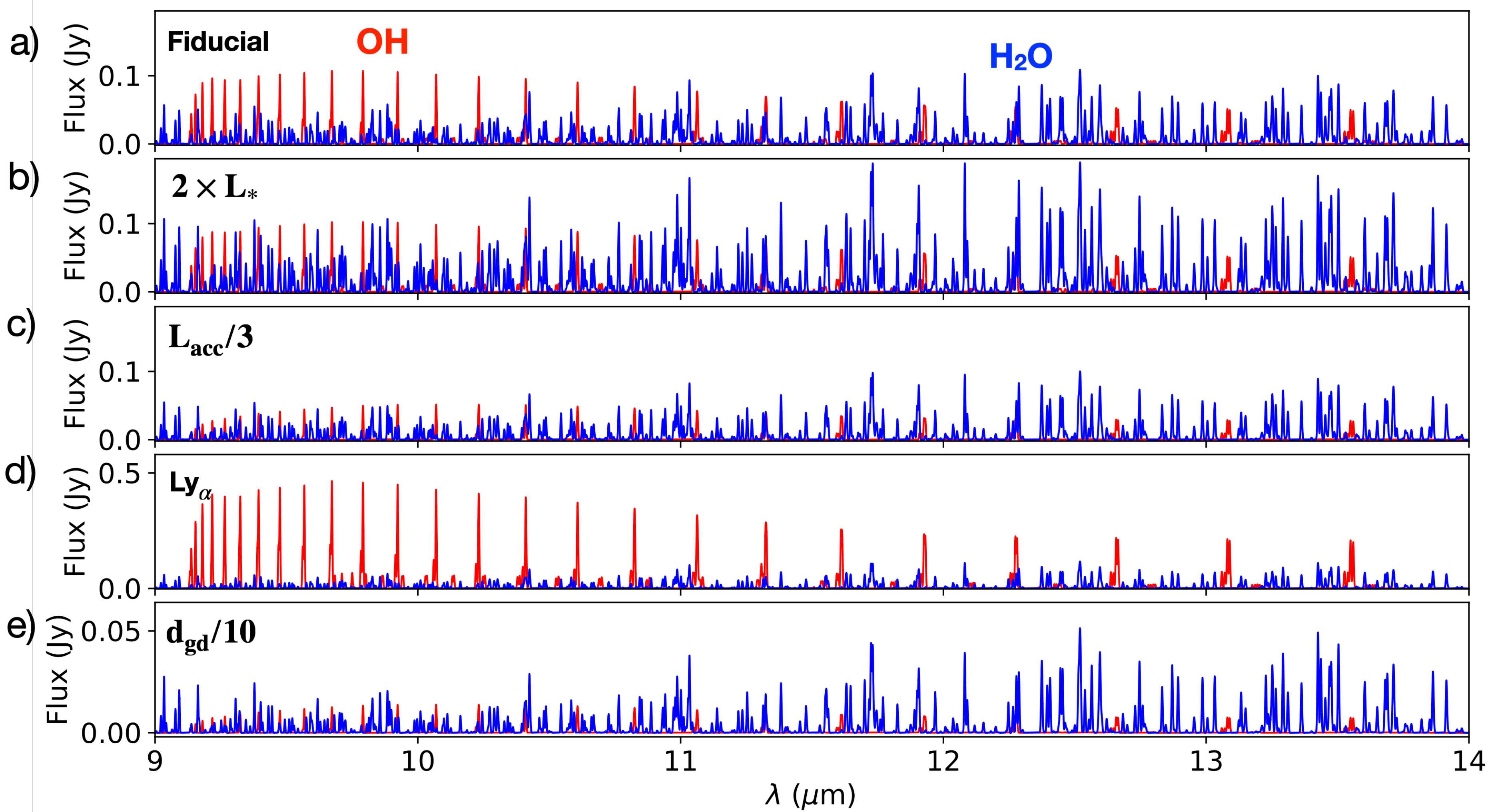}
\caption{Synthetic spectrum of \water and OH emission depending on the luminosity of the star, the accretion luminosity, the contribution of Ly$\alpha$ photons, and gas-to-dust mass ratio. We recall that the fiducial model is a disk with a gas-to-dust mass ratio of $d_{gd}=10^{5}$ orbiting a T Tauri star with a stellar luminosity of $L_*=1~L_{\odot}$, an accretion luminosity of $L_{acc}=0.12~L_{\odot}$, and a no contribution of Ly$\alpha$ to the impinging FUV field. Note the different scales in panels d) and e) highlighting the strong impact of the putative Ly$\alpha$ photons and the gas-to-dust ratio.}
\label{fig:grid-spectra}
\end{figure*}


The implication of Eq. (\ref{eq:I_PhiB}) is that rotationally excited OH lines do not directly depend on the OH abundance but on the abundance of \water in the exposed layers. The distribution of gas-phase water is shown in Figure \ref{fig:dali-intro-physico-chem}. We recover the two main reservoirs of gas-phase water described by many authors though with differing definitions: (a) the water-rich reservoir within the snow line ($r \simeq 1-2$ au) and close to the midplane ($z/r<0.4$), corresponding to the bulk part of \water that is well shielded but difficult to observe due to dust and gas opacity effects \citep{2019ApJ...875...96N}, (b) the irradiated water reservoir ($G_0 \gtrsim 10^{4}$) where molecules are actively photodissociated and reformed on short timescales ($\lesssim$ a few days). We also recover that the radial extent of this reservoir corresponds to the region where  \water shielding operates \citep{2022ApJ...930L..26B}.

Using the example of the H$_2$O ($15_{2,13} - 14_{1,14}$) line at 12.519~\mum  ($E_{\text{up}} =4,100~$K, $A_{i,j} = 1.5~\rm{s^{-1}}$), Fig. \ref{fig:dali-intro-physico-chem} shows that the brightest \water mid-IR lines are tracing this thin irradiated layer and not the bulk part of the \water in the inner disk. As described by e.g., \citet{2015A&A...582A.105A} and \citet{2022ApJ...930L..26B}, the mid-IR spectrum of water is indeed dominated by optically thick lines that readily saturate above $N(\rm{H_2O}) \gtrsim 10^{18}$ cm$^{-2}$ and require high temperatures to be excited ($T_{\rm{K}} \gtrsim 400$~K). It means that the \water emitting region is generally inside of the \water snow surface. Since \water snow surface moves toward the star closer to the mid-plane, the radial extent of the \water emitting region and the mid-plane snowline, which is of interest for planet formation, are not unambiguously linked. In our fiducial model, the radial extent of the water-rich atmosphere is smaller than the mid-plane snowline (see Fig. \ref{fig:dali-intro-physico-chem}). \resub{We refer to \citet{2016ApJ...827..113N} for an extensive study of which region is traced by different H$_2$O lines, including the far-IR and sub-millimeter domain.}

Interestingly, OH mid-IR emission originates from the same layer, though at a somewhat higher altitude where \water abundance is lower than $\simeq 10^{-4}$. This feature is due to the fact that the line emissivity is proportional to the product of the \water density and the strength of the UV radiation field. At high altitudes, \water is fully dissociated and thus under-abundant, while near the mid-plane the UV radiation is reduced by extinction. In other words, the OH line traces the optimal region where \water is abundant and exposed to the UV. In practice, the emitting area covers strongly irradiated regions ranging from $G_0=10^{5}$ up to $10^{8}$ where water is relatively abundant.

The OH emission is also radially confined to the inner disk, within 0.5~au. In particular, the line intensity drops steeply with the distance to the star beyond $\simeq 1~$au (see Fig. \ref{fig:phi-fiducial}). This is directly related to the radial distribution of \water and to the strength of the UV radiation field (see black lines in Fig. \ref{fig:dali-intro-physico-chem}): further away from the star, the gas and dust temperature is too low to activate efficient gas-phase formation of \water (see Sec. \ref{subsubsec:physico-chem}) and the radiation field drops due to geometrical dilution. Interestingly, the steep drop in OH line intensity around 1~au corresponds to the location where \water UV self-shielding stops operating due to too low column density of \water.

\subsubsection{Key physico-chemical processes for OH mid-IR emission}
\label{subsubsec:physico-chem}

Because OH mid-IR line intensities are proportional to the amount of \water photodestroyed in the upper layer, they are very sensitive to the physical and chemical conditions in these regions. In order to highlight the key processes and understand the dependence of OH line intensity on disk and stellar properties (see next section), we shall dissect the upper layer of our fiducial model, where OH emission originates. 

Figure \ref{fig:fiducial-vertical-structure}-a and b show the chemical and thermal structure of the IR active layer.  \resub{Across this molecular layer, destruction by X-ray-generated molecular ions is negligible compared to FUV photodissociation \citep{2014ApJ...786..135A}.}  Both \water and OH form via the neutral-neutral formation pathway shown in Eq. \ref{eq:O-chemistry} \citep{2009ApJ...701..142G}
which is initiated by the formation of H$_2$. As seen in Fig. \ref{fig:fiducial-vertical-structure}-a, it is indeed at the H/H$_2$ transition that \water starts to be abundant. Therefore, the density and the radiation field at the H/H$_2$ transition are of prime importance for the intensity of OH mid-IR lines. 

The H/H$_2$ transition is set by the balance between the formation and destruction of H$_2$. In the warm IR active layer, H$_2$ formation is primarily due to surface chemistry even though 3-body reactions tend to take over close to the star and for high gas-to-dust mass ratio ($r \lesssim 0.2~$au, $d_{gd} \gtrsim 10^4$). Interestingly, we find that H$_2$ is well self-shielded in the atomic layer ($z/r \lesssim 0.52$) and H$_2$ is primarily destroyed via the two-step process:
\begin{equation}
    \text{H}_2 + \text{O} \ce{<=>[ ][ ]} \text{OH} + \text{H} \ce{->[UV]} \text{O} + 2 \text{H}.
\end{equation}
Therefore, the presence of OH in the H/H$_2$ transition is the signpost of H$_2$ indirect photodestruction (see Fig. \ref{fig:fiducial-vertical-structure}-a). This constitutes a key difference with more diffuse PDRs at the edge of molecular clouds for which H$_2$ self-shielding, supplemented by dust attenuation, controls the H/H$_2$ transition \citep[e.g.,][]{1986ApJS...62..109V}. In that regard, the progressive increase in H$_2$ with decreasing altitude is driven by the decrease in far-UV flux due to dust attenuation (see Fig. \ref{fig:fiducial-vertical-structure}-a and b). One already notes that for the same gas distribution, the position of the H/H$_2$ transition will therefore depend on the abundance of small grains, which controls the attenuation of the UV field and the formation rate of H$_2$. The full conversion of H into H$_2$ at $z/r \simeq 0.44$ is also driven by H$_2$O UV shielding that reduces dramatically OH photodissociation and therefore the indirect destruction of H$_2$ (see Figure \ref{fig:fiducial-vertical-structure}-b, black versus grey lines).

Abundant \water in the irradiated layer also requires relatively high temperature because $ \text{H}_2 + \text{O} \rightarrow \text{OH} + \text{H} $ and $ \text{H}_2 + \text{OH} \rightarrow \text{H}_2\text{O} + \text{H} $ have high activation energies. Figure \ref{fig:fiducial-vertical-structure}-b shows that the gas is thermally coupled to the dust at H/H$_2$ with a temperature of about $600$~K, enough to lead to abundant \water despite the strong UV field ($x$(\water) $\simeq 10^{-5}$). The temperature-dependent formation of \water also controls the radial extent of the irradiated \water reservoir (Fig. \ref{fig:dali-intro-physico-chem}). As mentioned by \citet{2014ApJ...786..135A} and further discussed in \citet{2022ApJ...930L..26B}, the heating of the gas is dominated by OH and \water photodissociation \resub{with negligible contribution from X-ray heating}. We further stress that the inclusion of ro-vibrational cooling by CO and H$_2$O lines is particularly important in this layer as it brings the gas temperature close to that of the dust. \resub{Aditionally, vertical mixing, which is not included in the present model, can inject H$_2$ and H$_2$O to even more exposed layers, possibly enhancing the OH mid-IR emission \citep{2022A&A...668A.164W}.}

In summary, the amount of \water exposed to the strong UV field is deeply entangled with the physical conditions in the IR active layers via complex feedbacks: the H$_2$ abundance that sets the location of the IR active layer depends on its destruction by O followed by OH photodissociation. Following the increase in H$_2$ abundance, the \water abundance rises, forming the irradiated reservoir required for the mid-IR emission of OH. This means that OH mid-IR lines, in concert with \water lines, are promising diagnostics to test thermochemical disk models and constrain dust properties in the IR active layers.

\subsection{Grid of models}

\label{subsec:result-grid}

\begin{figure}
\centering
\includegraphics[width=0.5\textwidth]{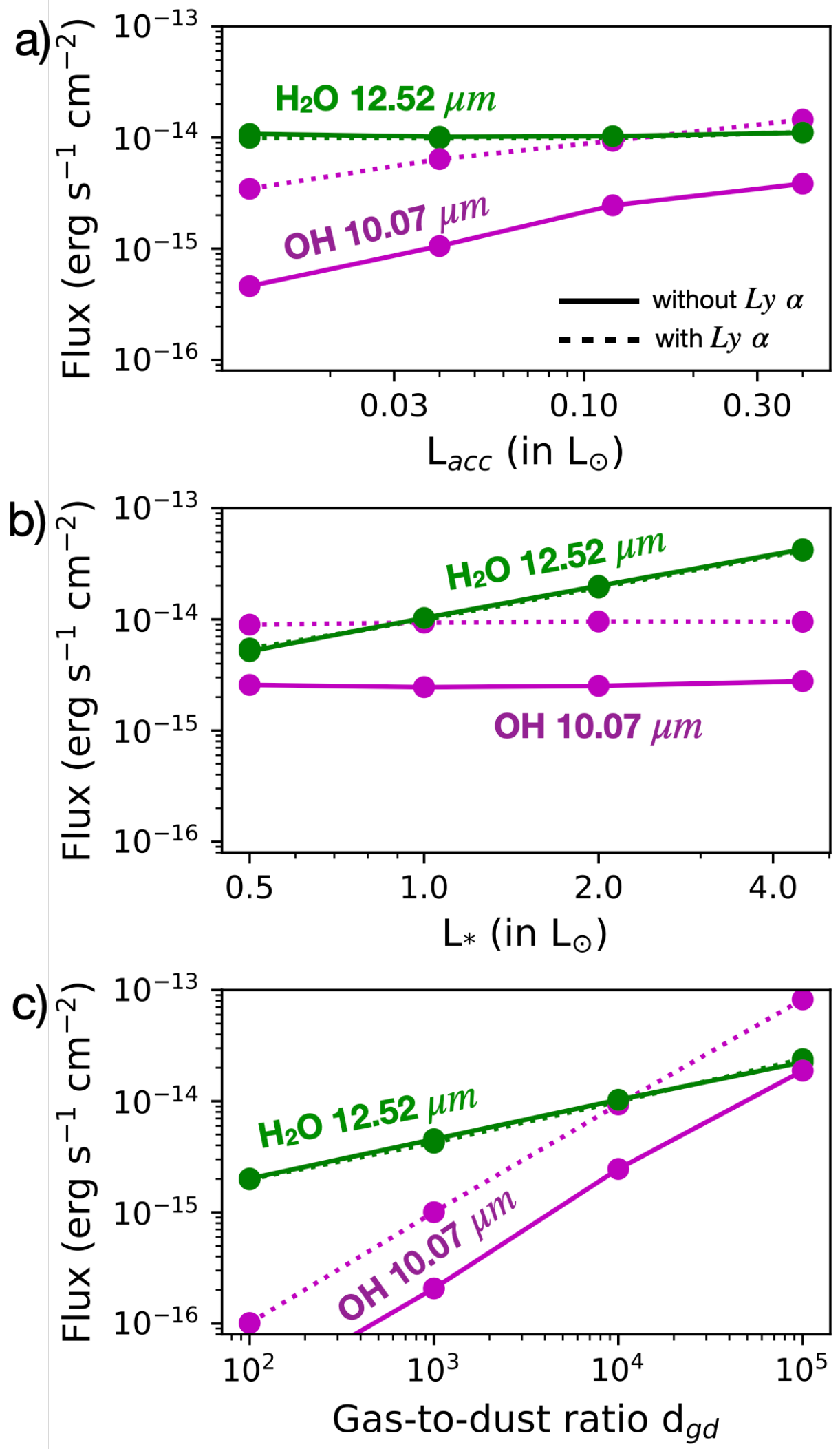}
\caption{Effect of the stellar and disk parameters on the OH and \water line fluxes. OH is primarily sensitive to the stellar FUV flux, which depends on $L_{acc}$, on the inclusion of Ly$\alpha$ photons, and on the gas-to-dust ratio. \water line flux is sensitive to the bolometric luminosity of the star and the gas-to-dust ratio. \resub{In this figure, the reference value of the parameters that are kept constant are $d_{gd}=10^4$, $L_*=1L_{\odot}$, and $L_{acc}=0.12 L_{\odot}$.}}
\label{fig:grid}
\end{figure}
Figure \ref{fig:grid-spectra} illustrates how \water and OH lines vary when changing key parameters, namely the luminosity of the star $L_*$, the accretion luminosity $L_{acc}$, the inclusion of Ly$\alpha$ photons, and the gas-to-dust mass ratio $d_{gd}$. OH emission appears to be sensitive to all of these parameters but $L_*$ whereas \water emission primarily depends on $L_*$ and on the gas-to-dust mass ratio  $d_{gd}$. These essential features are further explored on a large grid of models in Figure \ref{fig:grid} where we present the flux of the OH and H$_2$O blends at 10.06~\mum and 12.6$~\mu$m, respectively. These blends correspond to the $N=35 \rightarrow 34$ OH quadruplet and five \water lines dominated by upper energy levels of $\simeq 4,000$~K. We stress that for OH lines at longer wavelengths ($\gtrsim 13.5~\mu$m) these trends might be different as OH lines can be excited by chemical pumping.


\subsubsection{Stellar luminosity}

For fiducial disk parameters, we find that increasing the source luminosity $L_*$  does not impact the OH line flux but enhances the \water line flux (Fig. \ref{fig:grid-spectra}-b, and purple line Fig. \ref{fig:grid}-b). The fact that the OH line flux is independent of $L_{*}$ could be surprising since higher luminosity results in a globally warmer disk and therefore more efficient H$_2$O formation. In fact, OH emission is dominated by the inner disk, typically within $~0.8$~au. Changing the stellar luminosity has only a limited impact on the amount of H$_2$O in that specific region and therefore on the local line intensity (see radial intensity profiles in Fig. \ref{fig:appendix-intensity-profiles-Lstar}). Still, increasing the bolometric luminosity increases the radial extent of the H$_2$O rich exposed layer. This produces more extended OH emission but this contributes significantly to the total line flux only for $L_* \gtrsim 4.5 L_{\odot}$.

In contrast, the 12.5~\mum \water blend is thermally excited and optically thick across its emitting region. Increasing the bolometric luminosity has therefore two impacts illustrated in the Appendix (see Fig. \ref{fig:appendix-intensity-profiles-Lstar}). First, in the inner disk ($\lesssim$ 1 au), the temperature of the emitting layer is higher, and therefore the local intensity is higher ($I_{\nu} = B_{\nu}(T_K)$). Second, the region where dust and therefore gas is warm enough to sustain efficient \water formation in the irradiated layer extends to a larger distance. Therefore, the \water line flux increases with $L_{*}$ because the local intensity is higher and the emitting size increases. Quantitatively, for our fiducial disk parameters, the \water line flux is found to scale as $L_{bol}^{0.5}$, where $L_{bol}=L_*+L_{acc}$ is the bolometric luminosity of the star.

\subsubsection{Accretion luminosity}


OH being produced by H$_2$O photodissociation, the increase in line flux with increasing accretion luminosity is expected as the incident FUV flux is increased accordingly. However, this simple interpretation hides a more subtle balance in the physico-chemical processes. The OH emissivity is indeed proportional to the product $G_0 n_{\rm{H_2O}}$. Stronger FUV fields could then translate to less H$_2$O, canceling the expected increase in OH fluxes. In fact, as the incident FUV flux increases, the H/H$_2$ transition moves deeper into the disk where $n_{\text{H}}/G_0 \simeq 10^6-10^7~\rm{cm^{-3}}$ such that the gas density is higher. This allows \water to still form efficiently at the H/H$_2$ transition despite the increase in the incident UV field. The net balance is positive and OH line flux increases with $L_{acc}$. For our fiducial model, we find that  the OH flux scales as $\propto L_{acc}^{0.5}$. Because the gas and dust temperature is relatively constant in the vertical direction, this small change in emitting height does not impact the emission of the optically thick \water lines.



\subsubsection{Ly$\alpha$}
The OH line flux also depends strongly on the contribution of photons shortward of 144~nm to the accretion luminosity. When Ly$\alpha$ is added to the UV continuum excess, the OH line fluxes are increased by about a factor 4 over a wide range of $L_{acc}$ and $L_*$ (see dashed and solid purple lines in Fig. \ref{fig:grid}-a and b). In contrast, the \water line fluxes are not affected by Ly$\alpha$ photons.

The effect of including Ly$\alpha$ photons for a fixed value of $L_{acc}$ is to increase the amount of dissociating photons ($\lambda \lesssim 200$ nm). Therefore including Ly$\alpha$ photons for a fixed value of $L_{acc}$ has a similar effect as increasing $L_{acc}$. The analysis conducted above for the impact of $L_{acc}$ can be repeated: including Ly$\alpha$ photons results in a larger amount of H$_2$O photodissociated whereas H$_2$O formation is enhanced by the H/H$_2$ transition being deeper (higher densities). \water line fluxes are not affected because they depend primarily on the bolometric luminosity and less on the exact spectral distribution of the stellar spectrum.

\subsubsection{Gas-to-dust mass ratio}
Figure \ref{fig:grid}-c shows that OH line flux is approximately proportional to the gas-to-dust ratio whereas the increase in \water line flux is shallower ($\propto d_{gd}^{0.5}$). Therefore the gas-to-dust ratio is the only parameter that impacts both \water and OH lines. This is a result of a complex combination of thermal and chemical effects.

Depleting dust grains has three direct effects: (1) an increase in the penetration depth of the UV photons, (2) a reduction in H$_2$ formation rate on dust, and (3) an increase in dust temperature. The former two effects lead to a shift of the H/H$_2$ transition deeper into the disk where densities are larger keeping a high density of \water at high UV flux. Similar to an increase in $L_{acc}$, this leads to an increase in the amount of \water photodissociated and therefore of the OH lines. The increase in \water lines is mostly driven by the increase in dust and therefore gas temperature. Similar to the effect of increasing $L_{acc}$, the local line intensity of optically thick \water lines is larger and the emission increases at a larger distance. This results in an increase in \water line flux with the $d_{gd}$ ratio.



\section{Discussion}

\label{sec:discussion}

Our grid of DALI models shows that OH emission depends strongly on the disk parameters and is therefore a promising diagnostic to analyze the physical and chemical structure of IR active layers. Furthermore, \water and OH line fluxes exhibit distinct variation with basic disk parameters that allows disentangling the effects of different parameters. In this section, we intend to discuss our results and provide a first comparison with \textit{Spitzer}-IRS observations in preparation for further in-depth analysis of JWST-MIRI spectra where OH quadruplets are now seen in multiple disks, though at relatively long wavelengths \citep[$\lambda > 14~\mu$m][]{2023ApJ...947L...6G,2023ApJ...945L...7K,2023arXiv230709301G}.

\subsection{Measuring the local UV field}

We found that rotationally excited OH lines are optically thin over a large fraction of its emitting area and we recovered the results of \citetalias{2021A&A...650A.192T} that their fluxes are directly proportional to the amount of \water photodestroyed by photons in the 114-144~nm range. In \citetalias{2021A&A...650A.192T}, we proposed to infer the local FUV flux by estimating the photodissociation rate of \water [s$^{-1}$] as
\begin{equation}
    k_{B} = \Phi_{B}/N(\rm{H_2O}).
    \label{eq:kb}
\end{equation}
This method would be particularly promising for disks since it would give access to the FUV flux at the H/H$_2$ transition, a parameter that is key for all the disk chemical models and yet unconstrained. In this section, we discuss the caveats behind the applicability of Eq. (\ref{eq:kb}) and postpone the design of a model-independent approach to observationally infer the FUV flux in the disks' upper layers to a next paper.

One can already note that OH emission is radially extended and covers two orders of magnitude in $G_0$ (see Fig. \ref{fig:fiducial-vertical-structure}-b). This means that the value of $G_0$ estimated from Eq. (\ref{eq:kb}) is expected to be representative of a radius where the bulk part of OH emission originates from, which is $0.5~$au in our fiducial model. In a model-independent approach, measuring the radial extent of OH emission from spatially resolved observations appears unfeasible with JWST-MIRI and even with the ELT-METIS. One of the most promising methods is then to obtain spectrally resolved spectra of OH mid-IR lines, for example using the TEXES spectrograph \citep{2019ApJ...874...24S}. 

The use of Eq. (\ref{eq:kb}) also requires determining the column density of \water in the emitting region of OH only. In order to determine the \water column density, the mid-IR lines of \water could appear promising as they trace a similar radial extent. However, the OH emission is also confined to a very thin layer above the bulk part of \water reservoir. Therefore, the main challenge is to estimate the \water column density exposed to the FUV field. One could argue that the brightest \water lines are tracing an altitude close to that of OH (see the example of the \water line at 12.5 \mum in Fig. \ref{fig:fiducial-vertical-structure}-b). However, these lines are also optically thick, which limits our ability to directly infer the column density in the irradiated layer. 

A lower limit on the intensity of the FUV radiation field in the emitting layer of OH can still be placed. The attenuation of the radiation field is indeed due to a combination of dust attenuation and \water UV shielding. When \water shielding starts to operate, the FUV field drops steeply with altitude (see Fig. \ref{fig:fiducial-vertical-structure}, middle panel). Therefore, OH mid-IR emission comes necessarily from a layer with $N(\rm{H_2O}) \lesssim 10^{18}~$cm$^{-2}$. Injecting this upper limit into Eq. (\ref{eq:kb}) provides a lower limit on $k_{B}$ and therefore on the UV flux in the 114-144~nm range.

\subsection{Detectability with JWST/MIRI-MRS}

\begin{figure}
\centering
\includegraphics[width=0.5\textwidth]{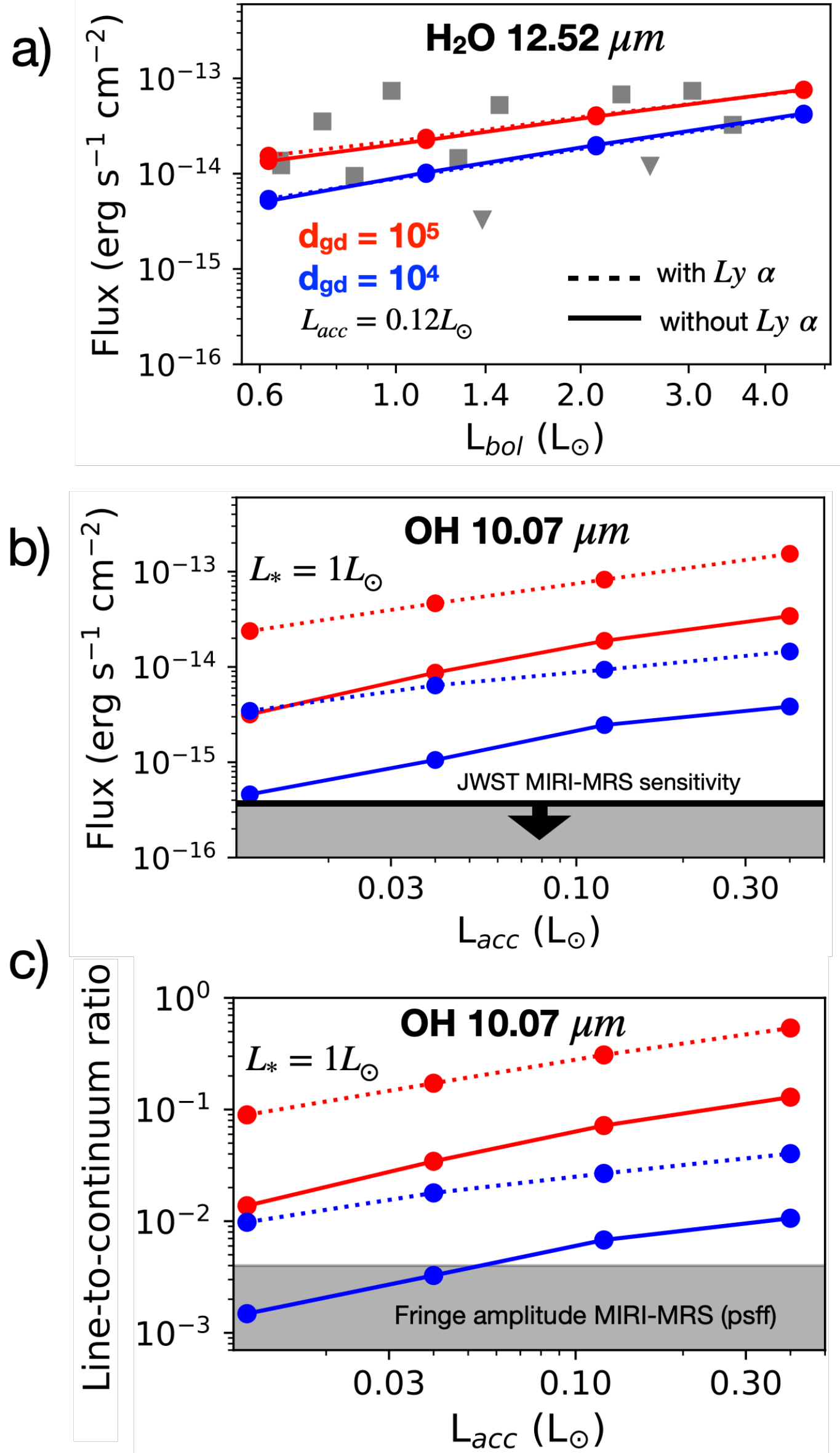}
\caption{Predictions for JWST MIRI-MRS. a) Comparison between \textit{Spitzer}-IRS fluxes of the \water blend at 12.5~$\mu$m and our DALI models for two values of gas-to-dust mass ratio. \textit{Spitzer}-IRS fluxes, represented by grey markers (triangles for upper limits and squares for detections), are well matched by dust-depleted disk atmospheres. The accretion luminosity is set to its fiducial value of $L_{acc}=0.12~L_{\odot}$, recalling that $L_{acc}$ has little effect on H$_2$O line fluxes. b) OH line fluxes at 10.07~$\mu$m as a function of $L_{acc}$. For dust depletion factors in line with \textit{Spitzer}-IRS water line fluxes and without considering MIRI-MRS fringes, OH is expected to be detected. c) The contrast between the OH line at 10.07~$\mu$m and the continuum. The grey area shows the typical residual noise after a \texttt{psff} fringe correction \citep{gasman_2023a} \resub{or calibration with observations of asteroids \citep{2024ApJ...963..158P}.}}
\label{fig:spitzer-vs-DALI}
\end{figure}

\BT{Several JWST MIRI spectra show OH lines in the 15~$\mu$m region \citep{2023ApJ...947L...6G,2023ApJ...945L...7K,2023arXiv230709301G} but no studies have yet been carried out at shorter wavelengths where OH emission can be uniquely attributed to prompt emission. In order to estimate the detectability of rotationally excited lines with MIRI-MRS, we apply our modeling results to previous \textit{Spitzer}-IRS results. Rotationally excited lines of OH shortward of $13~\mu$m have been analyzed in detail only in two disks: TW Hya, a transition disk, and DG Tau, a very strong accretor. In order to discuss the detectability of OH lines for the bulk population of T Tauri disks, we therefore focus on the H$_2$O line fluxes measured with \textit{Spitzer}-IRS to validate our DALI models and predict OH lines observable with JWST/MIRI-MRS.}

Our sample stems from the compilation of mid-IR \water line fluxes published by \citet{2017ApJ...834..152B}.  We further reduce the sample by selecting disks around T Tauri stars with an upper threshold in stellar mass of $1.4 ~M_{\odot}$ and a bolometric luminosity of $L_{bol}=5~L_{\odot}$. Since we focus on full disks, we excluded the disks with gas cavities by selecting the sources with an inner CO disk smaller than $R_{CO}=0.2~$au \citep{2015ApJ...809..167B}. This criterion is thought to be a better tracer of the inner gaseous disk than the infrared excess $n_{13-30}$ used by e.g. \citet{2007ApJ...664L.107B} \citep[see discussion in][Appendix D]{2020ApJ...903..124B}. With these criteria, we are left with a sample of 12 T Tauri disks with stellar luminosity ranging from 0.4 to 3.4~$L_{\odot}$ and accretion luminosity from 0.05 to 2~$L_{\odot}$. This allows us to explore the main trends and intrinsic scatter in the observed line fluxes. Of all the lines detected by \textit{Spitzer}-IRS, we select the 12.52 \mum \water blend integrated between 12.5 and 12.537~\mum for which we recall that our modeling predicts a similar emitting region as rotationally excited OH (see Fig. \ref{fig:dali-intro-physico-chem}). All the fluxes are rescaled to a distance of 140~pc.

Because our modeling predicts that the \water lines depend primarily on the bolometric luminosity of the star, we plot in Fig. \ref{fig:spitzer-vs-DALI}-a the measured 12.5~\mum \water line flux as a function of $L_{bol}$. The observations agree well with our prediction that \water line fluxes increase with $L_{bol}$. However, a large scatter in this correlation combined with the relatively narrow range in $L_{bol}$ prevent us from deriving robust scaling. The absolute line flux is well reproduced by models with a strong depletion of small grains in the atmosphere. In particular, the relatively large scatter in the $F_{\rm{H_2O}} - L_{bol}$ relation is well bracketed by our predictions with gas-to-dust mass ratio of $d_{gd} = 10^{4}-10^{5}$. This result is consistent with the early models of \citet{2009ApJ...704.1471M}, later confirmed by \citet{2015A&A...582A.105A} and \citet{2022ApJ...930L..26B}. An important caveat behind this comparison is that our sample is biased toward sources showing bright \water lines in the mid-IR. In fact, \water is detected in only 50\% of the T Tauri stars of G, K, and M type \citep{2010ApJ...720..887P}. It means that Fig. \ref{fig:spitzer-vs-DALI} represents only the tip of the distribution of \water flux. JWST observations will be key to have an unbiased sample with increased detection rates of \water. 

In contrast, our models predict that OH mid-IR lines are primarily sensitive to the FUV field reaching the warm molecular layer. Therefore, in Fig. \ref{fig:spitzer-vs-DALI}-b, we plot the observed OH flux as a function of the accretion luminosity $L_{acc}$ with and without Ly$\alpha$ photons and for the gas-to-dust ratios compatible with \textit{Spitzer}-IRS line fluxes. The stellar luminously is kept constant to its fiducial value ($L_* =1~L_{\odot}$). Interestingly, the OH line fluxes at 12.6~\mum measured by \textit{Spitzer}-IRS are well matched by our DALI models with a gas-to-dust ratio of $10^5$ (see Appendix \ref{fig:spitzer-vs-DALI_OH}). However, the \textit{Spitzer}-IRS line fluxes might be contaminated by the strong water lines in that region, and robust MIRI-MRS detection of these lines is required to refine this analysis. Here, we discuss the detectability of the OH lines with MRS-MRS. All the models predict line fluxes at 140~pc that are detectable by JWST-MIRI within 20~min of integration time \citep{2015PASP..127..686G} down to $L_{acc} \simeq 10^{-2} L_{\odot}$. As expected by the scalings found in this work, disks with a high gas-to-dust ratio and high accretion luminosity are prime targets to detect OH lines in the 9-13~\mum range. \resub{Interestingly, the FUV bump at 160~nm observed in T Tauri disks, when interpreted as a signature of water photodissociation via the channel O+H$_2$, indicates a typical amount of photodissociated water of about $10^{41}-10^{42}$ water molecules s$^{-1}$ \citep{2014ApJ...784..127F}. Converting this number into an OH mid-IR line flux (OH+H channel) we obtain typical values of about $10^{-14}-10^{-13}$ erg cm$^{-2}$ s$^{-1}$ (for a distance of 140~pc), in line with the range of predicted values shown in Fig. \ref{fig:spitzer-vs-DALI}.}  
We also note that for low gas-to-dust ratios and luminous stars, the contribution of H$_2$O lines is substantial even in the 10~\mum spectral region. A robust water line analysis, e.g. with 1D slab models predicting the entire forest of water lines, will be crucial to analyse the OH lines. 

\BT{One of the major limitations in the detection of OH lines by MIRI-MRS is spectral fringing, caused by coherent reflections inside the detector arrays \citep{2020A&A...641A.150A}. In Fig. \ref{fig:spitzer-vs-DALI}-c, we quantify the detection limit of the OH line at 10.07~\mum against residual fringe amplitude. The line-to-continuum ratio is found to depend strongly on the gas-to-dust ratio; a high gas-to-dust ratio results in a weaker continuum flux whereas the lines are enhanced. The possible contribution of Ly$\alpha$ increases substantially the line-to-continuum contrast by amplifying the OH lines. Quantitatively, fringe correction methods implemented in the standard JWST pipeline reduce the fringe amplitude to only about 3-10\% at 10~$\mu$m, which strongly limits the detectability of the lines. However, if the MIRI-MRS observation setup includes target acquisition and given the fact that OH emission is expected to be compact ($\lesssim 0.05"$) advanced fringe correction methods taking into account the dependence of the fringe properties on the MIRI-MRS pupil illumination and detector pixel sampling is recommended \citep[\texttt{psff} correction, see][]{gasman_2023a}. \resub{Alternatively, observations of asteroids as precise empirical reference sources allow to achieve high spectral contrast \citep{2024ApJ...963..158P}.} Typical residual fringe amplitudes are found to be about 0.5\%, enough to detect OH lines over a wide range of parameters (see grey area in Fig. \ref{fig:spitzer-vs-DALI}-c). We also note that our estimate of the line-to-continuum ratio is a pessimistic estimate since it is estimated at 10~\mum where the dust silicate feature is maximal.}

\subsection{FUV excess and the role of Ly$\alpha$}

The impact of Ly$\alpha$ photons onto the warm molecular layer constitutes a major uncertainty in all disk thermochemical models. The difficulty comes from the fact that Ly$\alpha$ photons, which represent up to 90\% of the FUV excess are scattered by H atoms, before being absorbed by dust or gas. Therefore the calculation of the propagation of Ly$\alpha$ photons requires coupling thermochemistry upon which the H abundance depends and radiative transfer. In our model, only dust scattering and absorption are taken into account.

\citet{2011ApJ...739...78B} first demonstrated that scattered Ly$\alpha$ photons can penetrate into the molecular layers, assuming that H$_2$ is formed on dust grains and destroyed by photodissociation. Our study of the H$_2$ chemistry (see Sec. \ref{subsec:result-fiducial}) shows however that, in inner disks, H$_2$ is also destroyed by atomic oxygen, resulting in a H/H$_2$ transition being closer to the mid-plane where the Ly$\alpha$ effective path-length is greatly increased. Including a complete chemical model and a simplified propagation model for Ly$\alpha$ photons, \citet{2016ApJ...817...82A} supports however the relevance of Ly$\alpha$ photons in the water-rich warm layer but using a low gas-to-dust ratio that is inconsistent with the observations of \water lines. Interestingly water photodissociation by Ly$\alpha$ photons is also evidenced by FUV observations \citep{2017ApJ...844..169F}. Yet, more recently, \citet{2022ApJ...934L..14C} used a realistic gas-to-dust ratio and an extensive chemical network and found that the Ly$\alpha$ photons have a negligible contribution to the FUV field in the molecular layers. All in all, the propagation of Ly$\alpha$ photons down to the molecular IR active layer remains debated. In this context, OH lines, in synergy with H$_2$O lines are promising avenues to constrain the propagation of Ly$\alpha$ photons down to the molecular layers.

In this work, we also modeled the FUV continuum excess as a blackbody at $20,000~$K. The exact shape of the FUV continuum excess remains uncertain because the observational constraints depend on the assumed dust extinction laws. In addition, a variation from source to source of the spectral slope is also found in UV observations \citep{2014ApJ...784..127F}. Contemporaneous observations of the FUV excess and of the mid-IR emission will be crucial in the JWST era. Our prediction of OH line flux as a function of the accretion luminosity should therefore be taken with caution. Whenever available, we recommend using the observational constraints of the FUV luminosity instead of $L_{acc}$, using the conversion factor assumed in our model of $L_{cont, FUV} = 0.03 L_{acc}$.


\subsection{Other parameters}

Throughout this work, we systematically explored the effect of a limited number of free parameters that are believed to be key. In this section, we briefly discuss the effect of the other parameters. Using a fixed dust property, we demonstrated that the dust abundance in disk atmospheres regulates the IR emission of both \water and OH. In particular, high values of the gas-to-dust are inferred from \textit{Spitzer}-IRS observations, in line with the previous modeling of \water IR emission. This has however been established using a single dust composition and a single dust size distribution. The decisive impact of dust is via the attenuation of the UV photons and the formation rate of H$_2$, two processes that involve the available dust surface per unit volume. Therefore, regardless of the exact dust size distribution, the key dust property is the effective cross-section per H atoms, that is, for our choice of dust distribution, $\sigma_{\rm{dust/H}} = 8 \times 10^{-22} (d_{\rm{gd}}/10^2)^{-1}$ cm$^2$. We also note that the dust size and composition affect dust temperature and therefore the \water lines. Conversely, our results show that the analysis of \water and OH lines, in concert with the continuum emission, is key for the understanding of dust evolution in inner disks \citep{2019A&A...626A...6G}. \resub{Abundant PAHs in the irradiated layers can enhance the formation of H$_2$ by PAH hydrogenation and H$_2$O formation via the increased photoelectric heating. These two effects increase the amount of photodissociated water and therefore the OH mid-IR line fluxes. We find that for the fiducial model, the abundance of PAH needs to have an abundance of at least $10^{-2}$ wrt ISM to affect the OH mid-IR lines. Synthetic predictions of PAH emission for specific UV radiation fields are required to adopt PAH abundances consistent with the non-detection of aromatic infrared bands around most T Tauri stars \citep{2006A&A...459..545G}.}

The disk gas mass was kept constant whereas the gas mass of T Tauri disks inferred from sub-mm continuum flux typically spans more than an order of magnitude \citep{
2022arXiv220309930M}. However, for a fixed gas-to-dust ratio, only a small dependency of \water and OH line fluxes on disk mass is found (less than 50\% between $M_D = 3 \times 10^{-2}-3 \times 10^{-3} M_{\odot}$). This result seems in tension with \citet{2015A&A...582A.105A} who found a strong dependency of mid-IR \water lines flux on disk mass. However, in the latter study, dust settling is computed in a self-consistent way such that high disk mass corresponds to a stronger depletion of dust in the IR active upper layers.

Throughout this work, we also assumed a full T Tauri disk with a smooth distribution of gas down to $0.1~$au. Following the analysis of \citet{2017ApJ...834..152B}, one could expect that any cavity in the gas would quench OH mid-IR lines. Determining the critical gap size above which OH emission would be quenched is beyond the scope of this paper. Interestingly, the disk of TW Hya, which has a gap of 2.4 au, exhibits a deficit in hot H$_2$O and OH near-IR lines but has detected OH and H$_2$O lines longward of 10 \mum \citep{2010ApJ...712..274N, 2024arXiv240309210H}. Moreover, the OH/H$_2$O flux ratio at 12.5 \mum is relatively high compared to the typical ratio found in full disks \citep[OH/H$_2$O=3 versus $\simeq 0.3$,][]{2017ApJ...834..152B}. This feature is likely due to the fact the 12.5 \mum H$_2$O blend needs high density and high temperature to be excited whereas OH can emit as long as there is enough \water to be photodissociated. Therefore, one can expect the OH/\water flux ratio 12.5 \mum to increase as a gap opens before the OH flux drops below the detection limit. For large gas cavities, OH emission is expected to be confined to the inner edge of the cavity where the FUV field directly illuminates the cavity wall, similar to the CO ro-vibrational emission in low NIR Group I Herbig disks \citep{2019A&A...631A.133B}. \resub{Disks with large cavities remain however an open topic as evidenced by the recent discovery of hot water in the disk of PDS 70 revealing active oxygen chemistry even in gas-depleted inner disks \citep{2023Natur.620..516P}. } \resub{We can also speculate that the presence of a forming gas-giant planet can locally increase the gas temperature \citep[e.g.,][]{2015ApJ...807....2C}, releasing water in the gas phase. The UV radiation emitted by such an accreting planet \citep{2018ApJ...866...84A} could then lead to a bright spot seen in OH mid-IR emission.}

\section{Conclusion}
\label{sec:conclusion}

In this work, we explore the potential of the rotationally excited OH lines in the mid-IR for the study of planet-forming disks. To reach this goal results from quantum mechanical calculations resolving the quantum state of the OH product following H$_2$O photodissociation are implemented in the DALI thermo-chemical disk model. An important addition compared to the modelling work of \citet{2021A&A...650A.192T} and \citet{2023A&A...671A..41Z} is that OH is assumed to be formed in the A' symmetric state, in line with recent quantum calculation studies \citep{2015JChPh.142l4317Z}. An extensive grid of models is computed, exploring the effect of the stellar and disk parameters. Synthetic JWST/MIRI-MRS predictions are then provided and compared to the available \textit{Spitzer} data. Our conclusions are 
\begin{itemize}
    \item For a typical T Tauri disk, the rotationally excited lines of OH are dominated by prompt emission shortward of $\simeq 13$ \mum. Throughout the bulk part of the disk, their intensities are proportional to the column density of water photodissociated in the 114-144~\mum range. Longer wavelength OH lines could be affected by chemical pumping which is not included in this study.
    \item OH mid-IR emission traces a thin layer of gas typically located within 1~au and close to the H/H$_2$ transition. For our fiducial model, this layer lies above the region that is well shielded by \water. 
    \item OH and \water lines, when analyzed in concert, provide key diagnostics about the distribution of dust in the upper layers and the flux of the FUV in the 114-144~nm band. OH line fluxes increase with the gas-to-dust mass ratio and the amount of FUV photons reaching the molecular layers. In particular, we show that they can constrain the amount of Ly$\alpha$ photons reaching the molecular layers, a parameter that remains debated. Focusing on the \water line blend at 12.5 \mum, we also find that \water increases with the gas-to-dust ratio and the bolometric luminosity of the star.
    \item Our model is able to reproduce \water emission at 12.5 \mum measured with \textit{Spitzer}-IRS with high gas-to-dust mass ratio. Based on these constraints, we expect OH lines to be detected by MIRI-MRS if Ly$\alpha$ photons reach the molecular layer, or for sources with high FUV continuum luminosity. Accurate fringe correction of the MIRI-MRS data is crucial to detect these lines on top of the bright dust continuum.
\end{itemize}

JWST, with its unique sensitivity and good spectral resolution, will be able to systematically probe the OH and \water emission in T Tauri disks as demonstrated by initial data. These data should also be supplemented by ground-based observations able to spectrally resolve the lines and measure the emitting area of OH. Detailed models including OH prompt emission will then allow constraining key disk parameters such as the gas-to-dust ratio and the UV field reaching the molecular layers. These constraints will be pivotal to infer elemental abundances from the observed molecular features lying in the IR. 
\newline
\newline

\noindent
\textit{\textbf{Data Availability.}} The synthetic spectra will be made available on a Zenodo repository along with the published version.
\newline

\begin{acknowledgements}
This work is part of the research programme Dutch Astrochemistry Network II with project number 614.001.751, which is (partly) financed by the Dutch Research Council (NWO). B.T. acknowledges the Paris Region fellowship program, which is supported by the Ile-de-France Region and has received funding under the Horizon 2020 innovation framework program and Marie Sklodowska-Curie grant agreement No. 945298. Part of this work was supported by the Programme National “Physique et Chimie du Milieu Interstellaire” (PCMI) of CNRS/INSU with INC/INP co-funded by CEA and CNES. E.v.D. acknowledges the funding from the European Research Council (ERC) under the European Union's Horizon 2020 research and innovation program (grant agreement No. 291141 MOLDISK).
\end{acknowledgements}

\bibliographystyle{aa} 
\bibliography{export-bibtex.bib} 

\begin{appendix}

\section{Intensity profiles}
\label{app:distrib}

\begin{figure*}
\centering
\includegraphics[width=1.0\textwidth]{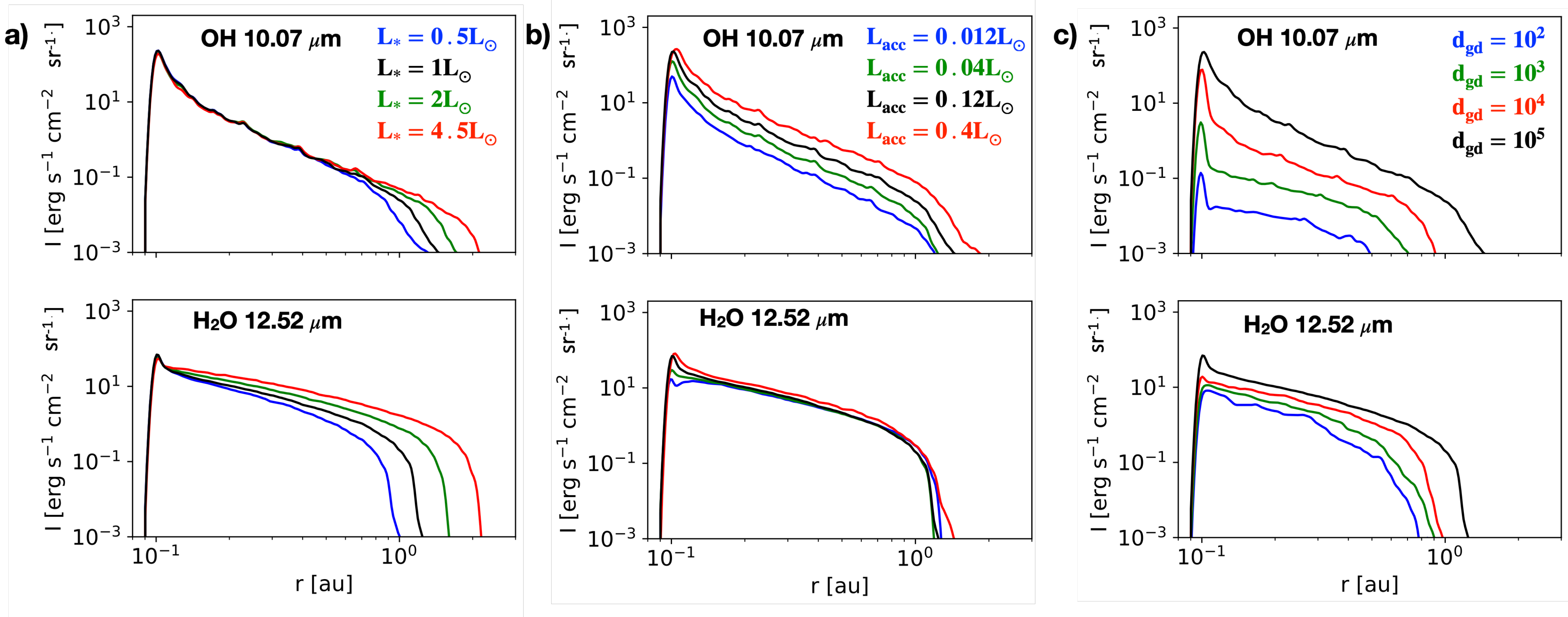}
\caption{Intensity profiles of the OH line at 10.07 \mum (top row) and the \water line at 12.52 \mum (bottom row) depending on a) the luminosity of the star, b) the accretion luminosity, and c) the gas-to-dust mass ratio. The other parameters are fixed to their fiducial values (see Table \ref{table:grid-parameters}).}
\label{fig:appendix-intensity-profiles-Lstar}
\end{figure*}

\section{Comparision between \textit{Spitzer}-IRS and DALI models}
\label{app:OH_fluxes}

\begin{figure}
\centering
\includegraphics[width=0.5\textwidth]{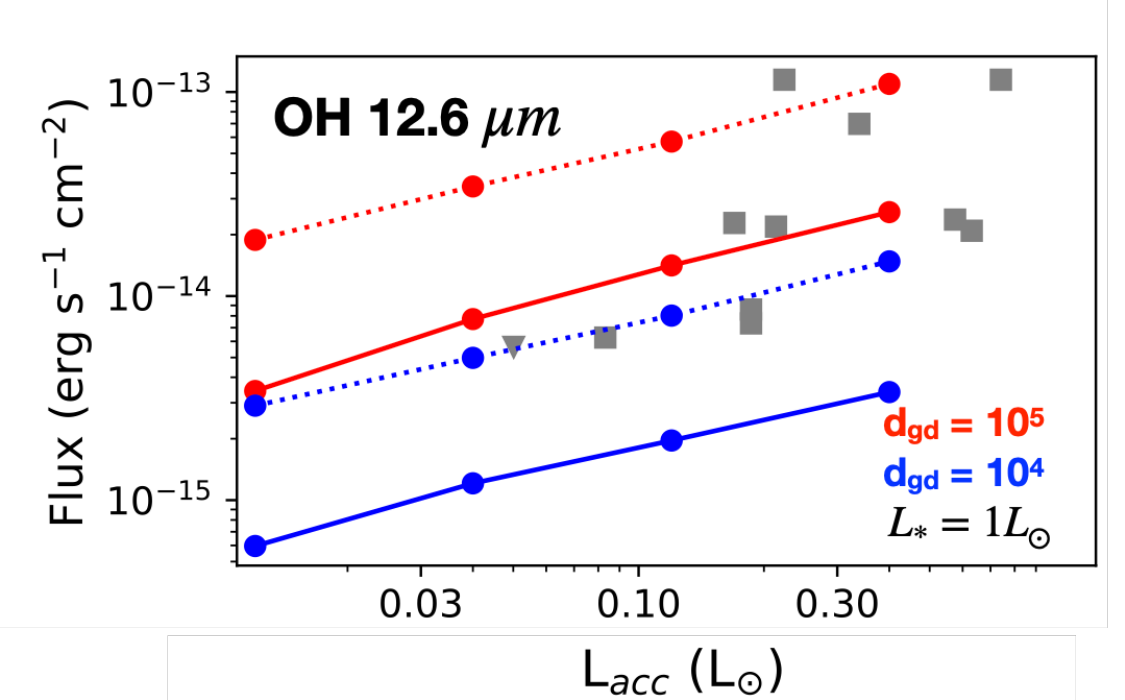}
\caption{Predictions of DALI model and \textit{Spitzer}-IRS flux of OH at 12.6~\mum. Spitzer-IRS fluxes, represented by grey makers (triangles for upper limits and squares for detections), are well matched by dust-depleted disk atmospheres.}
\label{fig:spitzer-vs-DALI_OH}
\end{figure}

\end{appendix}

\end{document}